\documentclass{article}

\usepackage{url}
\usepackage{amsmath}
\usepackage{graphicx}
\usepackage{epstopdf}
\usepackage{setspace}
\usepackage{geometry}
\usepackage{authblk}
\usepackage{fancyhdr}
\usepackage{color} 

\title{Bayesian Evidence Accumulation in Experimental Mathematics: A Case Study of Four Irrational Numbers}
\author{Quentin F. Gronau and Eric-Jan Wagenmakers}
\affil{University of Amsterdam}
\date{}

\begin{document}
\onehalfspacing
\maketitle
\begin{center}
Correspondence concerning this article should be addressed to:

Quentin F. Gronau

University of Amsterdam

Nieuwe Prinsengracht 130

1018 VZ Amsterdam, The Netherlands

E-mail may be sent to quentingronau@web.de

\end{center}

\begin{center}
\small
This work was supported by ERC grant 283876. Supplementary materials are available at \url{https://osf.io/5ysiu/}.
\end{center}

\begin{abstract}
Many questions in experimental mathematics are fundamentally inductive in nature. Here we demonstrate how Bayesian inference --the logic of partial beliefs-- can be used to quantify the evidence that finite data provide in favor of a general law. As a concrete example we focus on the general law which posits that certain fundamental constants (i.e., the irrational numbers $\pi$, $e$, $\sqrt{2}$, and $\ln{2}$) are normal; specifically, we consider the more restricted hypothesis that each digit in the constant's decimal expansion occurs equally often. Our analysis indicates that for each of the four constants, the evidence in favor of the general law is overwhelming. We argue that the Bayesian paradigm is particularly apt for applications in experimental mathematics, a field in which the plausibility of a general law is in need of constant revision in light of data sets whose size is increasing continually and indefinitely.
\end{abstract}

\section*{Introduction}

Experimental mathematics focuses on data and computation in order to address and discover mathematical questions that have so far escaped formal proof \cite{BaileyBorwein2009}. In many cases, this means that mathematical conjectures are examined by studying their consequences for a large range of data; every time a consequence is confirmed this increases one's confidence in the veracity of the conjecture. Complete confidence in the truth or falsehood of a conjecture can only be achieved with the help of a rigorous mathematical proof. Nevertheless, in between absolute truth and falsehood there exist partial beliefs, the intensity of which can be quantified using the rules of probability calculus \cite{Borel1965,Ramsey1926}.

Thus, an important role in experimental mathematics is played by heuristic reasoning and induction. Even in pure mathematics, inductive processes facilitate novel development:
\begin{quotation}
``every mathematician with some experience uses readily and effectively the same method that Euler used which is basically the following: To examine a theorem $T$, we deduce from it some easily verifiable consequences $C_1, C_2, C_3, \dotso$. If one of these consequences is found to be false, theorem $T$ is refuted and the question is decided. But if all the consequences $C_1, C_2, C_3, \dotso$ happen to be valid, we are led after a more or less lengthy sequence of verifications to an `inductive' conviction of the validity of theorem $T$. We attain a degree of belief so strong that it seems superfluous to make any ulterior verifications.'' \cite[pp. 455-456]{Polya1941}
\end{quotation}

Here we illustrate how to formalize the process of induction for a venerable problem in experimental mathematics: we will quantify degree of belief in the statement that particular irrational numbers (i.e., $\pi$, $e$, $\sqrt{2}$, and $\ln{2}$) are normal, or, more specifically, that the 10 digits of their decimal expansions occur equally often. This illustration does not address the more complicated question of whether all sequences of digits occur equally often: the sequence studied here is of length 1. Nevertheless, the simplified problem highlights the favorable properties of the general method and can be extended to more complicated scenarios.

To foreshadow the conclusion, our study shows that there is overwhelming evidence in favor of the general law that all digits in the decimal expansion of $\pi$, $e$, $\sqrt{2}$, and $\ln{2}$ occur equally often. Our statistical analysis improves on standard frequentist inference in several major ways that we elaborate upon below.

\section*{Bayes Factors to Quantify Evidence for General Laws}
In experimental mathematics, the topic of interest often concerns the possible existence of a general law. This law --sometimes termed the null hypothesis $\mathcal{H}_0$-- specifies an invariance (e.g., $\pi$ is normal) that imposes some sort of restriction on the data (e.g., the digits of the decimal expansion of $\pi$ occur equally often). The negation of the general law --sometimes termed the alternative hypothesis $\mathcal{H}_1$-- relaxes the restriction imposed by the general law.

In order to quantify the evidence that the data provide for or against a general law, \cite{Jeffreys1961} developed a formal system of statistical inference whose centerpiece is the following equation \sloppy{\cite[p. 387]{WrinchJeffreys1921}}:
\begin{equation}
\underbrace{\frac{p(\mathcal{H}_0 \mid \text{data})}{p(\mathcal{H}_1 \mid
\text{data})}}_{\text{Posterior
odds}}=\underbrace{\frac{p(\mathcal{H}_0)}{p(\mathcal{H}_1)}}_{\text{Prior
odds}} \times \,\,
\underbrace{\frac{p(\text{data} \mid \mathcal{H}_0)}{p(\text{data} \mid
\mathcal{H}_1)}}_{\text{Bayes factor BF}_{01}}.
\label{eq:postmodelratio}
\end{equation}
Jeffreys's work focused on the Bayes factor, which is the change from prior to posterior model odds brought about by the data. The Bayes factor also quantifies the relatively predictive adequacy of the models under consideration, and the log of the Bayes factor is the weight of evidence provided by the data \cite{KassRaftery1995}. When $\text{BF}_{01} = 10$ this indicates that the data are 10 times more likely under $\mathcal{H}_0$ than under $\mathcal{H}_1$; when $\text{BF}_{01} = .2$ this indicates that the data are 5 times more likely under $\mathcal{H}_1$ than under $\mathcal{H}_0$.

Let $\mathcal{H}_0$ be specified by a series of nuisance parameters $\zeta$ and, crucially, a parameter of interest that is fixed at a specific value, $\theta=\theta_0$. Then $\mathcal{H}_1$ is specified using similar nuisance parameters $\zeta$, but in addition $\mathcal{H}_1$ releases the restriction on $\theta$. In order to obtain the Bayes factor one needs to integrate out the model parameters as follows:
\begin{equation}
\text{BF}_{01} = \frac{\int_Z p(\text{data} \mid \theta_0, \zeta, \mathcal{H}_0) \, p(\zeta \mid \theta_0, \mathcal{H}_0)\, \text{d}\zeta}{\int_\Theta \int_Z p(\text{data} \mid \theta, \zeta, \mathcal{H}_1) \, p(\theta,\zeta \mid \mathcal{H}_1) \, \text{d}\zeta \, \text{d}\theta }.
\label{eq:BF}
\end{equation}

Equation~\ref{eq:BF} reveals several properties of Bayes factor inference that distinguish it from frequentist inference using $p$ values. First, the Bayes factor contrasts two hypotheses, the general law and its negation. Consequently, it is possible to quantify evidence in favor of the general law (i.e., whenever $\text{BF}_{01} > 1$). As we will see below, one of our tests for the first 100 million digits of $\pi$ produces $\mbox{BF}_{01}= 1.86\times10^{30}$, which is overwhelming evidence in favor of the law that the digits of the decimal expansion of $\pi$ occur equally often; in contrast, a non-significant $p$ value can only suggest a failure to reject $\mathcal{H}_0$ (e.g., \cite{Frey2009}). Moreover, as we will demonstrate below, the evidential meaning of a $p$ value changes with sample size \cite{Lindley1957}. This is particularly problematic for the study of the behavior of decimal expansions, since there can be as many as 10 trillion digits under consideration.

Second, the Bayes factor respects the probability calculus and allows coherent updating of beliefs; specifically, consider two batches of data, $y_1$ and $y_2$. Then, $\text{BF}_{01}(y_1,y_2) = \text{BF}_{01}(y_1) \times \text{BF}_{01}(y_2 \mid y_1)$: the Bayes factor for the joint data set can be decomposed as the product of the Bayes factor for the first batch multiplied by the Bayes factor for the second batch, conditional on the information obtained from the first data set. Consequently --and in contrast to $p$ value inference-- Bayes factors can be seamlessly updated as new data arrive, indefinitely and without a well-defined sampling plan \cite{BergerBerry1988a,BergerBerry1988b}. This property is particularly relevant for the study of normality of fundamental constants, since new computational and mathematical developments continually increase the length of the decimal expansion \cite{Wrench1960}.

\section*{The Normality of Irrational Numbers}
A real number $x$ is normal in base $b$ if all of the digit sequences in its base $b$ expansion occur equally often (e.g., \cite{Borel1909}); consequently, each string of $t$ consecutive digits has limiting frequency $b^{-t}$. In our example, we consider the decimal expansion and focus on strings of length 1. Hence, normality entails that each digit occurs with limiting frequency $1/10$.

The conjecture that certain fundamental constants --irrational numbers such as $\pi$, $e$, $\sqrt{2}$, and $\ln{2}$-- are normal has attracted much scientific scrutiny (e.g., \cite{BaileyBorwein2009,BaileyCrandall2001,BorweinBailey2004}). Aside from theoretical interest and practical application, the enduring fascination with this topic may be due in part to the paradoxical result that the digits sequences are perfectly predictable yet apparently appear random:
\begin{quotation}
``Plenty of arrangements in which design had a hand [...] would be quite indistinguishable in their results from those in which no design whatever could be traced. Perhaps the most striking case in point here is to be found in the arrangement of the digits in one of the natural arithmetical constants, such as $\pi$ or $e$, or in a table of logarithms. If we look to the process of production of these digits, no extremer instance can be found of what we mean by the antithesis of randomness: every figure has its necessarily pre-ordained position, and a moment's flagging of intention would defeat the whole purpose of the calculator. And yet, if we look to results only, no better instance can be found than one of these rows of digits if it were intended to illustrate what we practically understand by a chance arrangement of a number of objects. Each digit occurs approximately equally often, and this tendency developes [sic] as we advance further [...] In fact, if we were to take the whole row of hitherto calculated figures, cut off the first five as familiar to us all, and contemplate the rest, no one would have the slightest reason to suppose that these had not come out as the results of a die with ten equal faces.'' \cite[p. 111]{Venn1888}
\end{quotation}

But are constants such as $\pi$, $e$, $\sqrt{2}$, and $\ln{2}$ truly normal? Intuitive arguments suggest that normality must be the rule \cite[pp. 111-115]{Venn1888} but so far the problem has eluded a rigorous mathematical proof. In lieu of such a proof, research in experimental mathematics has developed a wide range of tests to assess whether or not the hypothesis of normality can be rejected (e.g., \cite{BaileyEtAl2012,Frey2009,Ganz2014,Jaditz2000,Marsaglia2005}; \cite[p. 281]{TuFischbach2005}), some of which involve visual methods of data presentation (e.g., \cite{ArtachoEtAl2012}; \cite[p. 118]{Venn1888}). In line with Venn's conjecture, most tests conclude that for the constants under investigation, the hypothesis of normality cannot be rejected.

However, to the best of our knowledge only one study has tried to quantify the strength of inductive support in favor of normality (i.e., \cite{BaileyEtAl2012}).
Below we outline a multinomial Bayes factor test of equivalence that allows one to quantify the evidence in favor of the general law that each digit occurs equally often.

\section*{A Bayes Factor Multinomial Test for Normality}
The general law or null hypothesis $\mathcal{H}_0$ states that $\pi$, $e$, $\sqrt{2}$, and $\ln{2}$ are normal. Here we consider the more restricted law that each digit in the decimal expansion occurs equally often (i.e., we focus on series of length 1 only). Hence, $\mathcal{H}_0$ stipulates that $\theta_{0j} = \frac{1}{10} \, \forall \, j \in \{0,1, \ldots ,9\}$, where $j$ indexes the digits.

Next we need to specify our expectations under $\mathcal{H}_1$, that is, our beliefs about the distribution of digit occurrences under the assumption that the general law does not hold, and before having seen actual data. We explore two alternative models. The first model assigns the digit probabilities $\theta_{j}$ an uninformative Dirichlet prior $D(\mathbf{a}=1)$; under this alternative hypothesis $\mathcal{H}_1^{\mathbf{a}=1}$, all combinations of digit probabilities are equally likely a priori. In other words, the predictions of $\mathcal{H}_1^{\mathbf{a}=1}$ are relatively imprecise. The second model assigns the digit probabilities $\theta_{j}$ an informative Dirichlet prior $D(\mathbf{a}=50)$; under this alternative hypothesis $\mathcal{H}_1^{\mathbf{a}=50}$, the predictions of $\mathcal{H}_1^{\mathbf{a}=50}$ are relatively precise, and similar to those made by $\mathcal{H}_0$. In effect, the predictions from $\mathcal{H}_1^{\mathbf{a}=50}$ are the same as those made by a model that is initialized with an uninformative Dirichlet prior $D(\mathbf{a}=1)$ which is then updated based on 49 hypothetical occurrences for each of the ten digits, that is, a hypothetical sequence of a total of 490 digits that corresponds perfectly with $\mathcal{H}_0$.

Thus, model $\mathcal{H}_1^{\mathbf{a}=1}$ yields predictions that are relatively imprecise, whereas model $\mathcal{H}_1^{\mathbf{a}=50}$ yields predictions that are relatively precise. The Bayes factor for $\mathcal{H}_0$ versus $\mathcal{H}_1$ is an indication of relative predictive adequacy, and by constructing two very different versions of $\mathcal{H}_1$ --one predictively dissimilar to $\mathcal{H}_0$, one predictively similar-- our analysis captures a wide range of plausible outcomes (e.g., \cite{SpiegelhalterEtAl1994}).

With $\mathcal{H}_0$ and $\mathcal{H}_1$ specified, the Bayes factor for the multinomial test of equivalence \cite[p. 350]{OHaganForster2004} is given by
\begin{equation} \label{BF01}
\begin{split}
\mbox{BF}_{01}  &= \frac{B(\mathbf{a})}{B(\mathbf{a}+\mathbf{n})}\prod_{j=0}^{9}\theta_{0j}^{n_j}\\
                &= \frac{B(\mathbf{a})}{B(\mathbf{a}+\mathbf{n})}\prod_{j=0}^{9}10^{-n_j},
\end{split}
\end{equation}
where $\mathbf{a}$ and $\mathbf{n}$ are vectors of length ten (i.e., the number of different digits); the elements of $\mathbf{n}$ contain the number of occurrences for each of the ten digits. Finally, $B(\cdot)$ is a generalization of the beta distribution \cite[p. 341]{OHaganForster2004}:
\begin{equation} \label{GenBeta}
B(\mathbf{a}) = \frac{\prod_{j=0}^{9}\Gamma(a_j)}{\Gamma\left(\sum_{j=0}^{9}a_j\right)},
\end{equation}
where $\Gamma(t)$ is the gamma function defined as $\Gamma (t)=\int _{0}^{\infty }x^{t-1}e^{-x}\,\text{d}x$. For computational convenience we use the natural logarithm of the Bayes factor:
\begin{equation} \label{logBF01}
\log\mbox{BF}_{01} = \log B(\mathbf{a})-\log B(\mathbf{a}+\mathbf{n})-N\log 10,
\end{equation}
where $N$ is the total number of observed digits.

\subsection*{Example 1: The Case of $\pi$}
\begin{figure}
\centering
\includegraphics[width= \textwidth]{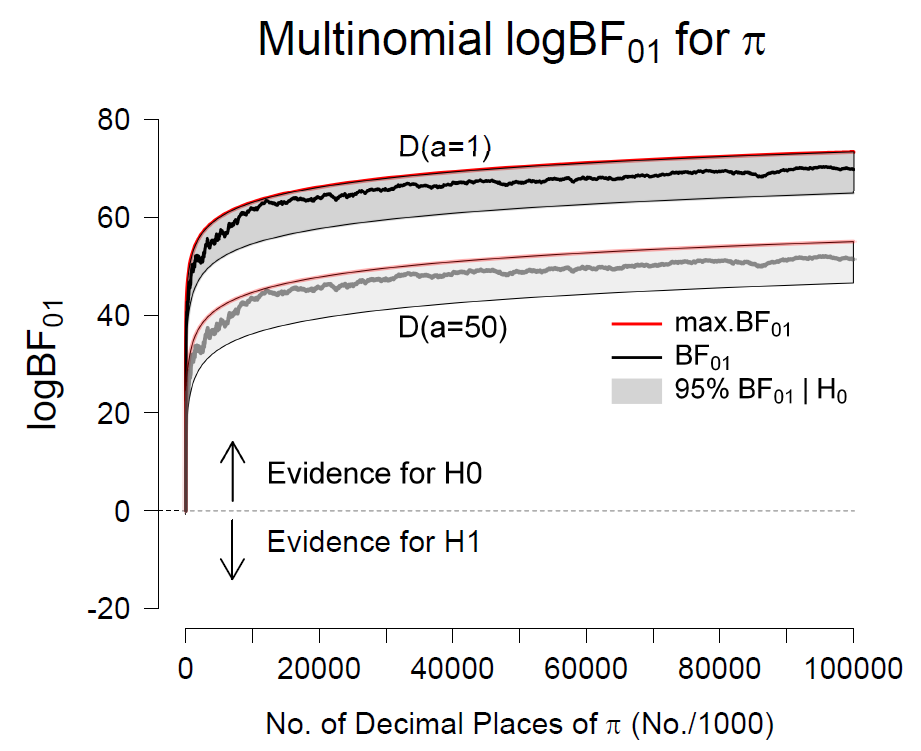}
\caption{Sequential Bayes factors in favor of equal occurrence probabilities based on the first 100 million digits of $\pi$.  The results in the top part of the panel correspond to an uninformative $D(\mathbf{a}=1)$ prior for the alternative hypothesis; the results in the lower part of the panel correspond to the use of an informative $D(\mathbf{a}=50)$ prior. The red lines indicate the maximum possible evidence for $\mathcal{H}_0$, and the grey areas indicate where 95\% of the Bayes factors would fall if $\mathcal{H}_0$ were true. After 100 million digits, the final Bayes factor under a $D(\mathbf{a}=1)$ prior is $\mbox{BF}_{01}= 1.86\times10^{30}$ ($\log \mbox{BF}_{01} = 69.70$); under a $D(\mathbf{a}=50)$ prior, the final Bayes factor equals $\mbox{BF}_{01}= 1.92\times10^{22}$ ($\log \mbox{BF}_{01} = 51.31$). Figure available at \protect \url{http://tinyurl.com/zelm4o4} under CC license \protect \url{https://creativecommons.org/licenses/by/2.0/}.}
\label{figure:pi}
\end{figure}

In our first example we compute multinomial Bayes factors for the digits of $\pi$. We compute the Bayes factor sequentially, as a function of an increasing number of available digits, with an upper bound of 100 million. Figure~\ref{figure:pi} displays the results in steps of 1,000 digits. The Bayes factor that contrasts $\mathcal{H}_0$ versus $\mathcal{H}_1^{\mathbf{a}=1}$ is indicated by the black line, and it shows that the evidence increasingly supports the general law. After all 100 million digits have been taken into account, the observed data are $1.86\times10^{30}$ times more likely to occur under $\mathcal{H}_0$ than under $\mathcal{H}_1^{\mathbf{a}=1}$. The extent of this support is overwhelming. The red line indicates the maximum Bayes factor, that is, the Bayes factor that is obtained in case the digits were to occur equally often -- that is, hypothetical data perfectly consistent with $\mathcal{H}_0$.

The dark grey area in Figure~\ref{figure:pi} indicates where a frequentist $p$ value hypothesis test would fail to reject the null hypothesis. This area was determined in two steps. First, we considered the hypothetical distribution of counts across the ten digit categories and constructed a threshold data set for which $\mathcal{H}_0$ has a 5\% chance of producing outcomes that are at least as extreme. Second, this threshold data set was used to compute a Bayes factor, and this threshold Bayes factor is plotted in Figure~\ref{figure:pi} as the lower bound of the dark grey area.

In order to construct the threshold data set, the number of counts in each digit category was obtained as follows. In this multinomial scenario there are nine degrees of freedom. Without loss of generality, the number of counts in the first eight of ten categories may be set equal to the expected frequency of $\frac{N}{10}$: $n_0, n_1,\ldots,n_7 = \frac{N}{10}$. Consequently, the first eight summands of the $\chi^2$-test formula are equal to zero. Furthermore, $\sum_{j=0}^{9}n_j=N$, so that if $n_8$ is known, $n_{9}$ is determined by $n_{9}=\frac{2}{10}N-n_8$. We then obtain the number of counts in the ninth category $n_8$ by solving the following quadratic equation for $n_8$:
\begin{equation}\label{n9}
\chi^2_{95 \%} = \frac{\left(n_8-\frac{N}{10}\right)^2}{N/10}+\frac{\left(\left(\frac{2}{10}\thinspace N-n_8\right)-\frac{N}{10}\right)^2}{N/10},
\end{equation}
where $\chi^2_{95 \%}$ denotes the $95$-th percentile of the $\chi^2$ distribution with nine degrees of freedom.

Figure~\ref{figure:pi} shows that the height of the dark grey area's lower bound increases with $N$. This means that it is possible to encounter a data set for which the Bayes factor indicates overwhelming evidence in favor of $\mathcal{H}_0$, whereas the fixed-$\alpha$ frequentist hypothesis test suggests that $\mathcal{H}_0$ ought to be rejected. In this way Figure~\ref{figure:pi} provides a visual illustration of the Jeffreys-Lindley paradox \cite{Jeffreys1961,Lindley1957}, a paradox that will turn out to be especially relevant for the later analysis of $e$, $\sqrt{2}$, and $\ln{2}$.

A qualitative similar pattern of results is apparent when we consider the grey line in Figure~\ref{figure:pi}: the Bayes factor that contrasts $\mathcal{H}_0$ versus $\mathcal{H}_1^{\mathbf{a}=50}$. Because this model makes predictions that are relatively similar to those of $\mathcal{H}_0$, the data are less diagnostic than before. Nevertheless, the evidence increasingly supports the general law. After all 100 million digits are observed, the observed data are $\mbox{BF}_{01}= 1.92\times10^{22}$ times more likely to occur under $\mathcal{H}_0$ than under $\mathcal{H}_1^{\mathbf{a}=50}$. The extent of this support remains overwhelming.

For completeness, we also computed Bayes factors based on the first trillion decimal digits of $\pi$ as reported in \cite[p. 11]{BaileyBorwein2009} (not shown). As expected from the upward evidential trajectories in Figure~\ref{figure:pi}, increasing the sequence length strengthens the support in favor of the general law: based on one trillion decimal digits,  the $D(\mathbf{a}=1)$ prior for $\mathcal{H}_1$ yields $\mbox{BF}_{01}= 3.65\times10^{46}$ ($\log \mbox{BF}_{01} = 107.29$)\footnote{Such an excessive degree of evidence in favor of a general law may well constitute a world record.}, and the $D(\mathbf{a}=50)$ prior yields $\mbox{BF}_{01}= 4.07\times10^{38}$ ($\log \mbox{BF}_{01} = 88.90$).

Finally, consider the fact that the two evidential trajectories --one for a comparison against $\mathcal{H}_1^{\mathbf{a}=1}$, one for a comparison against $\mathcal{H}_1^{\mathbf{a}=50}$-- have a similar shape and appear to differ only by a constant factor. This pattern is not a coincidence, and it follows from the nature of sequential updating for Bayes factors \cite[p. 334]{Jeffreys1961}. Recall that there exist two mathematically equivalent ways to update the Bayes factor when new data $y_2$ appear. The first method is to compute a single new Bayes factor using all of the available observations, $\text{BF}(y = y_1,y_2)$; the second method is to compute a Bayes factor only for the new data, but based on the posterior distribution that is the result of having encountered the previous data -- this Bayes factor, $\text{BF}(y_2 \mid y_1)$ is then multiplied by the Bayes factor for the old data, $\text{BF}(y_1)$ to yield the updated Bayes factor $\text{BF}(y = y_1,y_2)$.

Now let $y_1$ denote a starting sequence of digits large enough so that the joint posterior distribution for the $\theta_{j}$'s under $\mathcal{H}_1^{\mathbf{a}=1}$ is relatively similar to that under $\mathcal{H}_1^{\mathbf{a}=50}$ (i.e., when the data are said to have overwhelmed the prior). From that point onward, the change in the Bayes factor as a result of new data $y_2$, $\text{BF}(y_2 \mid y_1)$, will be virtually identical for both instantiations of $\mathcal{H}_1$. Hence, following an initial phase of posterior convergence, the subsequent evidential updates are almost completely independent of the prior distribution on the model parameters.\footnote{That is, after a sufficient number of observations, the trajectories of the log Bayes factors for the different priors for $\mathcal{H}_1$ are equal, only shifted by a constant.
In fact, regardless of the irrational number under consideration, this constant --which corresponds to the difference in $\log(\text{BF}_{01}^{\mathbf{a}=1})$ and $\log(\text{BF}_{01}^{\mathbf{a}=50})$-- approaches 18.39 (for a derivation see \url{https://osf.io/m5jas/}).}

Equation~\ref{eq:postmodelratio} shows that the Bayes factor quantifies the change in belief brought about by the data; as a first derivative of belief (expressed on the log scale), it achieves independence of the prior model log odds. In turn, Figure~\ref{figure:pi} illustrates that the change in the log Bayes factor --the second derivative of belief-- achieves independence of the prior distribution on the model parameters, albeit only in the limit of large samples.

The next three cases concern a study of the irrational numbers $e$, $\sqrt{2}$, and $\ln{2}$; the analysis and conclusion for these cases echo the ones for the case of $\pi$.

\subsection*{Example 2: The Case of $e$}
\begin{figure}
\centering
\includegraphics[width= \textwidth]{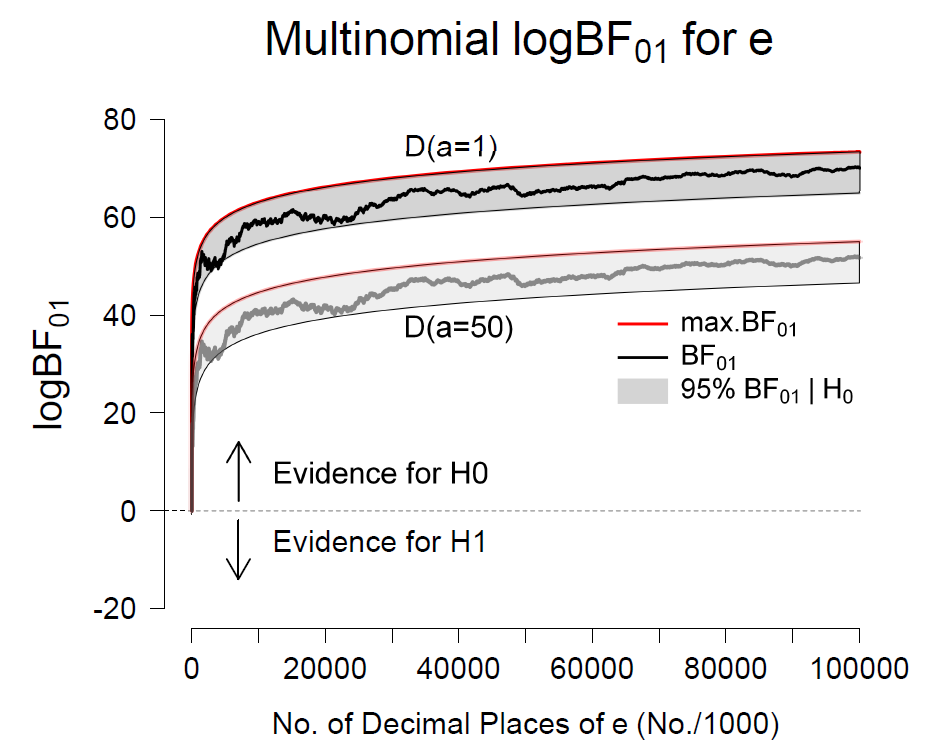}
\caption{Sequential Bayes factors in favor of equal occurrence probabilities based on the first 100 million digits of $e$.  The results in the top part of the panel correspond to an uninformative $D(\mathbf{a}=1)$ prior for the alternative hypothesis; the results in the lower part of the panel correspond to the use of an informative $D(\mathbf{a}=50)$ prior. The red lines indicate the maximum possible evidence for $\mathcal{H}_0$, and the grey areas indicate where 95\% of the Bayes factors would fall if $\mathcal{H}_0$ were true. After 100 million digits, the final Bayes factor under a $D(\mathbf{a}=1)$ prior is $\mbox{BF}_{01}= 2.61\times10^{30}$ ($\log \mbox{BF}_{01} = 70.04$); under a $D(\mathbf{a}=50)$ prior, the final Bayes factor equals $\mbox{BF}_{01}= 2.69\times10^{22}$ ($\log \mbox{BF}_{01} = 51.65$). Figure available at \protect \url{http://tinyurl.com/h3wenqo} under CC license \protect \url{https://creativecommons.org/licenses/by/2.0/}.}
\label{figure:e}
\end{figure}

In our second example we compute multinomial Bayes factors for the digits of the base of the natural logaritm: Euler's number $e$. Proceeding in similar fashion as for the case of $\pi$, Figure~\ref{figure:e} shows the evidential trajectories (in steps of 1,000 digits) for the first 100 million digits of $e$.\footnote{Data were obtained using the \texttt{pifast} software (\url{	numbers.computation.free.fr/Constants/PiProgram/pifast.html}).} As was the case for $\pi$, the upward trajectories signal an increasing degree of support in favor of the general law. After all 100 million digits have been taken into account, the observed data are $2.61\times10^{30}$ times more likely to occur under $\mathcal{H}_0$ than under $\mathcal{H}_1^{\mathbf{a}=1}$, and $2.69\times10^{22}$ times more likely under $\mathcal{H}_0$ than under $\mathcal{H}_1^{\mathbf{a}=50}$. Again, the extent of this support is overwhelming.

Note that, as for the case of $\pi$, the two evidential trajectories --one for a comparison against $\mathcal{H}_1^{\mathbf{a}=1}$, one for a comparison against $\mathcal{H}_1^{\mathbf{a}=50}$-- have a similar shape and appear to differ only by a constant factor. In contrast to the case of $\pi$, however, the Jeffreys-Lindley paradox is more than just a theoretical possibility: Figure~\ref{figure:e} shows that the evidential trajectories move outside the grey area when the total digit count is between $82,100$ and $254,000$, meaning that for those digit counts the frequentist hypothesis test (with a fixed $\alpha$-level of $.05$) suggests that $\mathcal{H}_0$ ought to be rejected. For the same data, both Bayes factors indicate compelling evidence in favor of $\mathcal{H}_0$.\footnote{A frequentist statistician may object that this is a sequential design whose proper analysis demands a correction of the $\alpha$ level. However, the same data may well occur in a fixed sample size design. In addition, the frequentist correction of $\alpha$ levels is undefined when the digit count increases indefinitely.}

\subsection*{Example 3: The Case of $\sqrt{2}$}
\begin{figure}
\centering
\includegraphics[width= \textwidth]{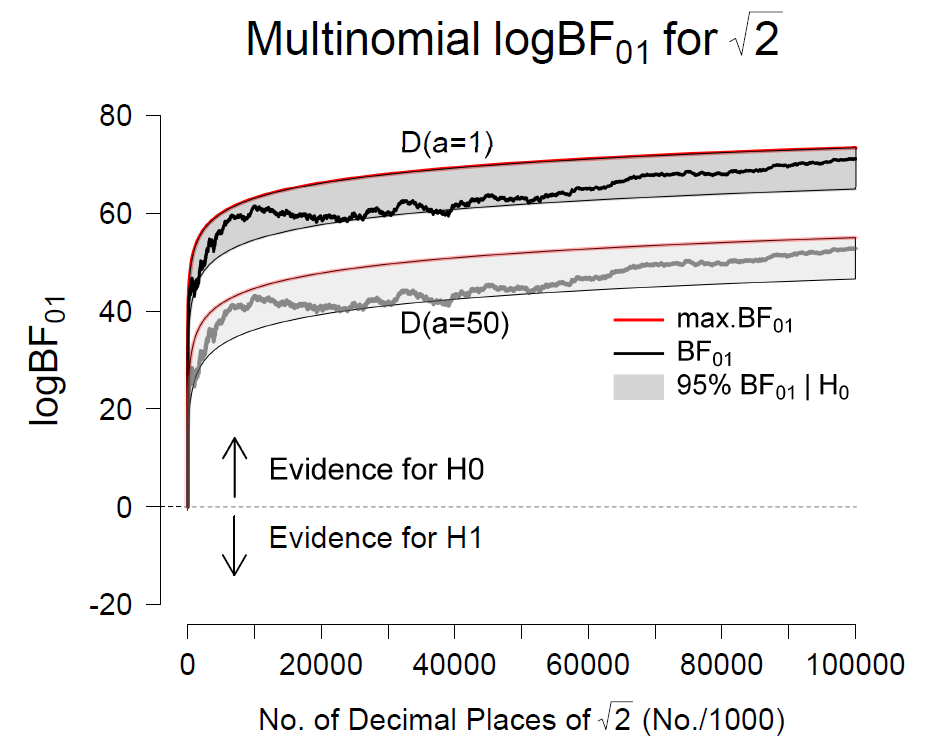}
\caption{Sequential Bayes factors in favor of equal occurrence probabilities based on the first 100 million digits of $\sqrt{2}$.  The results in the top part of the panel correspond to an uninformative $D(\mathbf{a}=1)$ prior for the alternative hypothesis; the results in the lower part of the panel correspond to the use of an informative $D(\mathbf{a}=50)$ prior. The red lines indicate the maximum possible evidence for $\mathcal{H}_0$, and the grey areas indicate where 95\% of the Bayes factors would fall if $\mathcal{H}_0$ were true. After 100 million digits, the final Bayes factor under a $D(\mathbf{a}=1)$ prior is $\mbox{BF}_{01}= 7.29\times10^{30}$ ($\log \mbox{BF}_{01} = 71.06$); under a $D(\mathbf{a}=50)$ prior, the final Bayes factor equals $\mbox{BF}_{01}= 7.52\times10^{22}$ ($\log \mbox{BF}_{01} = 52.67$). Figure available at \protect \url{http://tinyurl.com/jgwu523} under CC license \protect \url{https://creativecommons.org/licenses/by/2.0/}.}
\label{figure:sqrt2}
\end{figure}

In our third example we compute multinomial Bayes factors for the digits of $\sqrt{2}$. Proceeding in similar fashion as above, Figure~\ref{figure:sqrt2} shows the evidential trajectories (in steps of 1,000 digits) for the first 100 million digits of $\sqrt{2}$.\footnote{Data were obtained using the \texttt{pifast} software (\url{	numbers.computation.free.fr/Constants/PiProgram/pifast.html}).} As was the case for $\pi$ and $e$, upward evidential trajectories reveal an increasing degree of support in favor of the general law. After all 100 million digits have been taken into account, the observed data are $7.29\times10^{30}$ times more likely to occur under $\mathcal{H}_0$ than under $\mathcal{H}_1^{\mathbf{a}=1}$, and $7.52\times10^{22}$ times more likely under $\mathcal{H}_0$ than under $\mathcal{H}_1^{\mathbf{a}=50}$. As before, the extent of this support is overwhelming.

As Figure~\ref{figure:sqrt2} shows, the analysis of $\sqrt{2}$ provides yet another demonstration of the Jeffreys-Lindley paradox: when the total digit count ranges between 1 million and 2 million, and between 20 and 40 million (especially close to 40 million), a frequentist analysis occasionally rejects $\mathcal{H}_0$ at an $\alpha$-level of $.05$ (i.e., the evidential trajectories temporarily leave the grey area) whereas, for the same data, both Bayes factors indicate compelling evidence in favor of $\mathcal{H}_0$.

\subsection*{Example 4: The Case of $\ln{2}$}
\begin{figure}
\centering
\includegraphics[width= \textwidth]{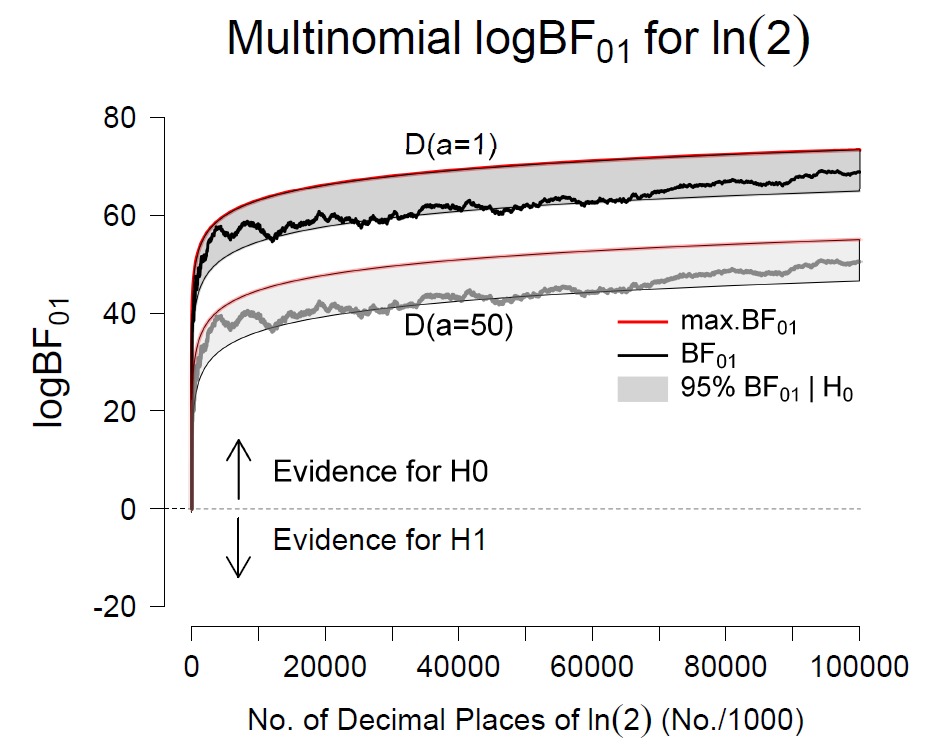}
\caption{Sequential Bayes factors in favor of equal occurrence probabilities based on the first 100 million digits of $\ln{2}$.  The results in the top part of the panel correspond to an uninformative $D(\mathbf{a}=1)$ prior for the alternative hypothesis; the results in the lower part of the panel correspond to the use of an informative $D(\mathbf{a}=50)$ prior. The red lines indicate the maximum possible evidence for $\mathcal{H}_0$, and the grey areas indicate where 95\% of the Bayes factors would fall if $\mathcal{H}_0$ were true. After 100 million digits, the final Bayes factor under a $D(\mathbf{a}=1)$ prior is $\mbox{BF}_{01}= 7.58\times10^{29}$ ($\log \mbox{BF}_{01} = 68.80$); under a $D(\mathbf{a}=50)$ prior, the final Bayes factor equals $\mbox{BF}_{01}= 7.81\times10^{21}$ ($\log \mbox{BF}_{01} = 50.41$). Figure available at \protect \url{http://tinyurl.com/jqdyd3w	} under CC license
\protect \url{https://creativecommons.org/licenses/by/2.0/}.}
\label{figure:ln2}
\end{figure}

In our fourth and final example we compute multinomial Bayes factors for the digits of $\ln{2}$. Figure~\ref{figure:ln2} shows the evidential trajectories (in steps of 1,000 digits) for the first 100 million digits of $\ln{2}$.\footnote{Data were obtained using the \texttt{pifast} software (\url{	numbers.computation.free.fr/Constants/PiProgram/pifast.html}).} As was the case for $\pi$, $e$, and $\sqrt{2}$, upward trajectories reflect the increasing degree of support in favor of the general law. After all 100 million digits have been taken into account, the observed data are $7.58\times10^{29}$ times more likely to occur under $\mathcal{H}_0$ than under $\mathcal{H}_1^{\mathbf{a}=1}$, and $7.81\times10^{21}$ times more likely under $\mathcal{H}_0$ than under $\mathcal{H}_1^{\mathbf{a}=50}$. As Figure~\ref{figure:ln2} shows, the analysis of $\ln{2}$ provides again a demonstration of the Jeffreys-Lindley paradox: the evidential trajectories leave the grey area multiple times indicating that a frequentist analysis rejects $\mathcal{H}_0$ at an $\alpha$-level of $.05$ whereas, for the same data, both Bayes factors indicate compelling evidence in favor of $\mathcal{H}_0$.

\begin{figure}
	\centering
	\includegraphics[width= \textwidth]{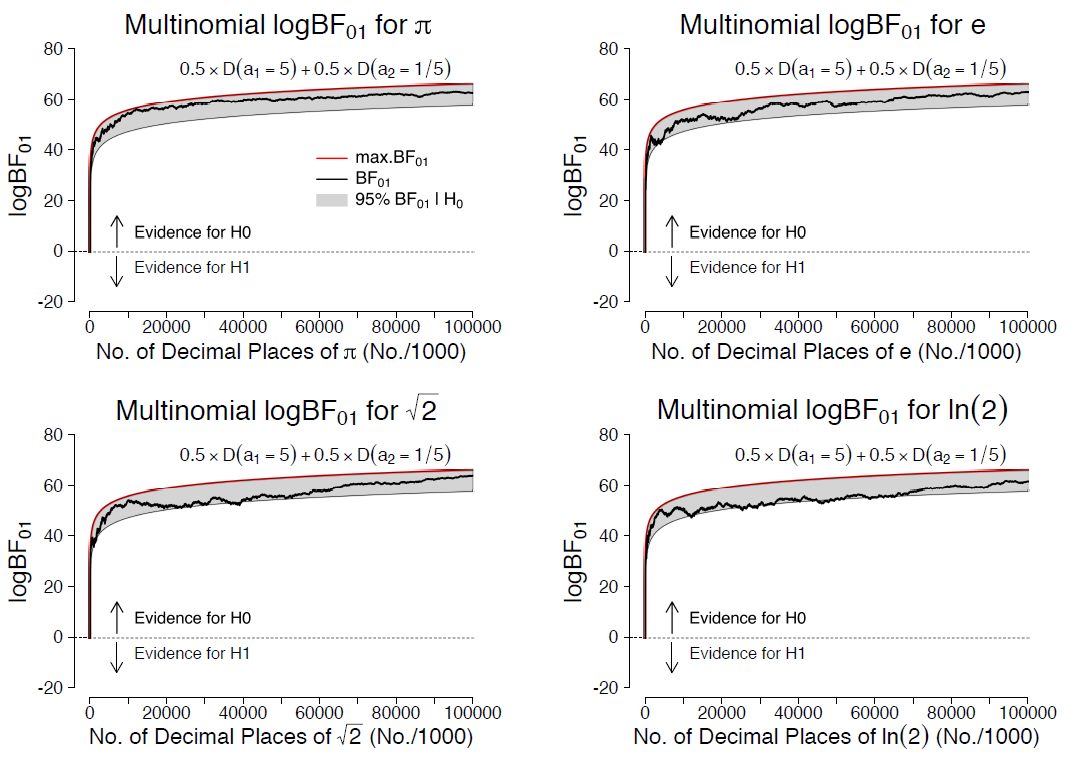}
	\caption{Sequential Bayes factors in favor of equal occurrence probabilities based on the first 100 million digits of $\pi$, $e$, $\sqrt{2}$, and $\ln{2}$.  The results correspond to the use of a two component mixture prior of a $D(\mathbf{a_1}=5)$ and $D(\mathbf{a_2}=1/5)$ Dirichlet distribution where the mixing weight was equal to $w=0.5$. The red lines indicate the maximum possible evidence for $\mathcal{H}_0$, and the grey areas indicate where 95\% of the Bayes factors would fall if $\mathcal{H}_0$ were true. Figure available at \protect \url{http://tinyurl.com/hw4gmlr} under CC license \protect \url{https://creativecommons.org/licenses/by/2.0/}.}
	\label{figure:alternative}
\end{figure}

\section*{Alternative Analysis}
The analyses presented so far used two different Dirichlet distributions as a prior for the parameter vector under the alternative hypothesis $\mathcal{H}_1$. 
In this way, we demonstrated that the results do not change qualitatively when considering an uninformed or an informed Dirichlet prior distribution.
A Dirichlet distribution is commonly used as a prior distribution for the parameter vector of a multinomial likelihood since it conveniently leads to an analytical solution for the Bayes factor.

However, one might ask whether the results are sensitive to the particular choice of the \emph{family} of prior distributions used to specify the alternative hypothesis $\mathcal{H}_1$, that is the family of Dirichlet distributions.
To highlight the robustness of our conlusion, we present the results of an analysis that is based on a more flexible prior distribution than the Dirichlet distribution, namely a two component mixture of Dirichlet distributions.
Mixture distributions have the property that the shape of the density is extremely flexible and can easily account for skewness, excess kurtosis, and even multi-modality \cite{FruehwirthSchnatter2006} which makes them an ideal candidate for testing the sensitivity to a wide range of prior distributions.
As \cite{DalalHall1983} showed, in fact \emph{any} prior distribution may be arbitrarily closely approximated by a suitable mixture of \emph{conjugate} prior distributions (i.e., prior distributions that, combined with a certain likelihood, lead to a posterior distribution that is in the same family of distributions as the prior distribution).\footnote{Of course, in some cases this may require a very ``rich'' mixture, that is, a mixture prior with many components.}

As an example, we considered a two component mixture of a $D(\mathbf{a_1}=5)$ Dirichlet distribution which assigns more mass to probability vectors that have components that are similar to each other (i.e., similar digit probabilities) and a $D(\mathbf{a_2}=1/5)$ Dirichlet distribution which assigns more mass to the corners of the simplex (i.e., one digit probability dominates) where the mixing weight was equal to $w = 0.5$.\footnote{\texttt{R} code that allows one to explore how the results change for a different choice of a two component Dirichlet mixture prior is available on the Open Science Framework under \url{https://osf.io/cmn2z/}.}
It is easily shown that also under this prior choice, the Bayes factor is available analytically.
Recall that the Bayes factor is defined as $\text{BF}_{01} = \frac{p(\text{data} \mid \mathcal{H}_0)}{p(\text{data} \mid \mathcal{H}_1)}$.  $p(\text{data} \mid \mathcal{H}_0)$ is obtained by inserting $\theta_{0j} = \frac{1}{10} \, \forall \, j \in \{0,1, \ldots,9\}$ into the multinomial likelihood.
In order to obtain $p(\text{data} \mid \mathcal{H}_1)$, we use the fact that any mixture of conjugate prior distributions is itself conjugate, that is, leads to a posterior distribution that is again a mixture of the same family of distributions, only with updated parameters \cite{DalalHall1983}.
Hence, since the Dirichlet distribution is conjugate to the multinomial likelihood, the posterior distribution when using a mixture of Dirichlet distributions as a prior is again a mixture of Dirichlet distributions (with updated parameters).
This implies that we know the normalizing constant of the posterior distribution under the alternative hypothesis $\mathcal{H}_1$ which is equivalent to $p(\text{data} \mid \mathcal{H}_1)$.
Hence, we can calculate the Bayes factor as follows:
\begin{equation}
\label{eq:mixtureBF}
\begin{split}
\text{BF}_{01} &= \frac{p(\text{data} \mid \mathcal{H}_0)}{p(\text{data} \mid \mathcal{H}_1)}\\
&=\frac{\frac{N!}{n_0!n_1!\ldots n_9!} \prod_{j = 0}^{9} \theta_{0j}^{n_j}}{\int\limits_{\mathbf{\Theta}}\frac{N!}{n_0!n_1!\ldots n_9!} \prod_{j = 0}^{9} \theta_j^{n_j} \big(w\thinspace \frac{1}{B(\mathbf{a_1})} \prod_{j=0}^{9} \theta_j^{a_{1j}-1} + (1-w) \thinspace \frac{1}{B(\mathbf{a_2})} \prod_{j=0}^{9} \theta_j^{a_{2j}-1}\big) \text{d}\mathbf{\theta}}\\
&= \frac{\prod_{j= 0}^{9} \theta_{0j}^{n_j}}{w \thinspace \frac{1}{B(\mathbf{a_1})}\int\limits_{\mathbf{\Theta}}\prod_{j=0}^{9} \theta_j^{a_{1j}+n_j-1}\text{d}\mathbf{\theta} + (1-w) \thinspace \frac{1}{B(\mathbf{a_2})}\int\limits_{\mathbf{\Theta}} \prod_{j=0}^{9} \theta_j^{a_{2j}+n_j-1} \text{d}\mathbf{\theta}} \\
&= \frac{\prod_{j = 0}^{9} \theta_{0j}^{n_j}}{w\thinspace \frac{B(\mathbf{a_1} + \mathbf{n})}{B(\mathbf{a_1})} + (1-w) \thinspace \frac{B(\mathbf{a_2} + \mathbf{n})}{B(\mathbf{a_2})}}.
\end{split}
\end{equation}

Figure~\ref{figure:alternative} displays the results for the 100 million digits of the four irrational numbers that are based on the two component mixture prior described above.
For $\pi$, the final Bayes factor equals $1.41\times10^{27}$; for $e$, the final Bayes factor equals $1.97\times10^{27}$; for $\sqrt{2}$, the final Bayes factor equals $5.52\times10^{27}$; for $\ln2$ the final Bayes factor equals $5.73\times10^{26}$.

The results based on the mixture prior are very similar to the previous ones, that is, we again obtain overwhelming support in favor of the assumption that all digits occur equally often; hence, we conclude that inference appears to be relatively robust to the particular choice of prior distribution that is used.

\begin{figure}
	\centering
	\includegraphics{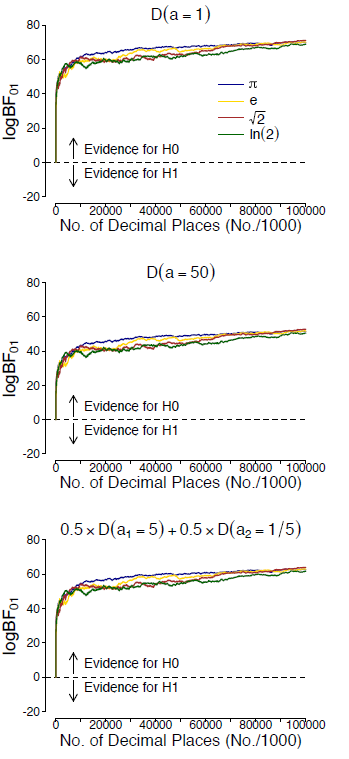}
	\caption{Sequential Bayes factors in favor of equal occurrence probabilities based on the first 100 million digits of  $\pi$, $e$, $\sqrt{2}$, and $\ln{2}$.  The results in the upper panel correspond to the use of an uninformative $D(\mathbf{a}=1)$ prior for the alternative hypothesis; the results in the middle panel correspond to the use of an informative $D(\mathbf{a}=50)$ prior; the results in the lower panel correspond to the use of a two component mixture prior of a $D(\mathbf{a_1}=5)$ and $D(\mathbf{a_2}=1/5)$ Dirichlet distribution where the mixing weight was equal to $w=0.5$. Figure available at \protect \url{http://tinyurl.com/hhut8dp} under CC license \protect \url{https://creativecommons.org/licenses/by/2.0/}.}
	\label{figure:3panel}
\end{figure}

\section*{Discussion and Conclusion}
With the help of four examples we illustrated how Bayesian inference can be used to quantify evidence in favor of a general law \cite{Jeffreys1961}. Specifically, we examined the degree to which the data support the conjecture that the digits in the decimal expansion of $\pi$, $e$, $\sqrt{2}$, and $\ln{2}$ occur equally often.
Our main analysis featured two prior distributions used to instantiate models as alternatives to the general law: the alternative model $\mathcal{H}_1^{\mathbf{a}=50}$ resembled the general law, whereas the alternative model $\mathcal{H}_1^{\mathbf{a}=1}$ did not. An infinite number of plausible alternatives and associated inferences lie in between these two extremes. Regardless of whether the comparison involved $\mathcal{H}_1^{\mathbf{a}=50}$ or $\mathcal{H}_1^{\mathbf{a}=1}$, the evidence was always compelling and the sequential analysis produced evidential trajectories that reflected increasing support in favor of the general law. Future data can update the evidence and extend these trajectories indefinitely.

Figures~\ref{figure:pi}--\ref{figure:ln2} clearly show the different outcomes for $\mathcal{H}_1^{\mathbf{a}=50}$ versus $\mathcal{H}_1^{\mathbf{a}=1}$. This dependence on the model specification is sometimes felt to be a weakness of the Bayesian approach, as the specification of the prior distribution for the model parameters is not always straightforward or objective. However, the dependence on the prior distribution is also a strength, as it allows the researcher to insert relevant information into the model to devise a test that more closely represents the underlying theory. Does it make sense to assign the model parameters a Dirichlet $D(\mathbf{a}=50)$ prior? It is easy to use existing knowledge about the distribution of trillions of digits for $\pi$ to argue that this Dirichlet distribution is overly wide and hence inappropriate; however, this conclusion confuses prior knowledge with posterior knowledge -- as the name implies, the prior distribution should reflect our opinion before and not after the data have been observed.

In the present work we tried to alleviate concerns about the sensitivity to the prior specification in three ways. First, for our main analysis, we used a sandwich approach in which we examined the results for two very different prior distributions, thereby capturing a wide range of outcomes for alternative specifications (e.g., \cite{SpiegelhalterEtAl1994}).
Second, we considered a different, very flexible family of alternative prior distributions (i.e., a two component mixture of Dirichlet distributions) and we demonstrated that the results do not change qualitatively -- the evidence in favor of the general law remains overwhelming.
Third, we have shown that the second derivative of belief --the change in the Bayes factor as a result of new data-- becomes insensitive to the prior specification as $N$ grows large. Here, the evidential trajectories all suggest that the evidence for the general law increases as more digits become available.
Figure~\ref{figure:3panel} displays the results for $\pi$, $e$, $\sqrt{2}$, and $\ln{2}$ side-by-side and emphasizes that for all four irrational numbers that we investigated, we obtain similar overwhelming support for the general law which states that all digits occur equally often -- this is the case for all three prior distributions that we considered.

\begin{figure}
\centering
\includegraphics[width= \textwidth]{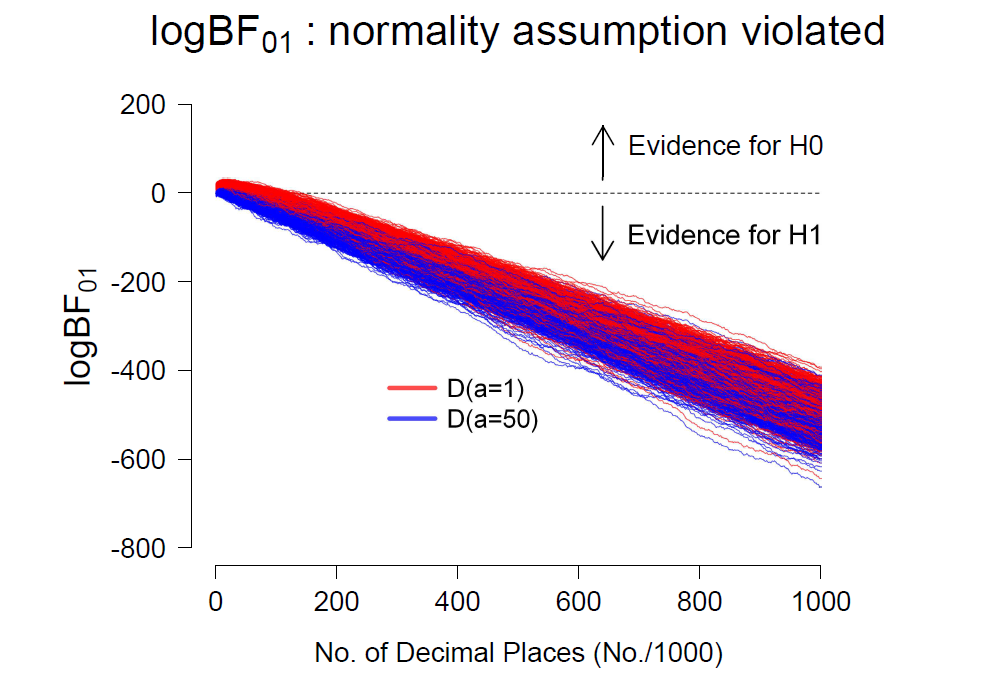}
\caption{Sequential Bayes factors in favor of equal occurrence probabilities for $1,000$ simulated data sets of 1 million digits each. In every data set, one digit was given an occurrence probability of $.11$ whereas each of the other digits occurred with probability $.89/9$. The evidential trajectories indicate increasingly strong evidence against the general law. Figure available at \protect \url{http://tinyurl.com/j4qk2ht} under CC license \protect \url{https://creativecommons.org/licenses/by/2.0/}.}
\label{figure:simulation}
\end{figure}

A remaining concern is that our Dirichlet prior on $\mathcal{H}_1^{\mathbf{a}=50}$ may be overly wide and therefore bias the test in favor of the general law. To assess the validity of this concern we conducted a simulation study in which the normality assumption was violated: one digit was given an occurrence probability of $.11$, whereas each of the remaining digits were given occurrence probabilities of $.89/9$. Figure~\ref{figure:simulation} shows that for all $1,000$ simulated data sets, the evidential trajectories indicate increasing evidence against the general law. After 1 million digits, the average Bayes factor in favor of the alternative hypothesis is $\mbox{BF}_{10}= 1.19\times10^{214}$ ($\log \mbox{BF}_{10} = 492.93$) under the $D(\mathbf{a}=1)$ prior and $\mbox{BF}_{10}= 8.88\times10^{221}$ ($\log \mbox{BF}_{10} = 511.05$) under the $D(\mathbf{a}=50)$ prior. Thus, with our instantiations of $\mathcal{H}_1$ the Bayes factor is able to provide overwhelming evidence against the general law when it is false.

One of the main challenges for Bayesian inference in the study of normality for fundamental constants is to extend the simple multinomial approach presented here to account for longer digit sequences. As the digit series grows large, the number of multinomial categories also grows while the number of unique sequences decreases. Ultimately, this means that even with trillions of digits, a test for normality may lack the data for a diagnostic test. Nevertheless, alternative models of randomness can be entertained and given a Bayesian implementation -- once this is done, the principles outlined by Jeffreys can be used to quantify the evidence for or against the general law.

\bibliographystyle{apalike}
\bibliography{referenties,referentiesQ}

\end{document}